\documentclass{amsart}

\usepackage{amsmath}
\usepackage{mathtools}
\usepackage{amssymb}
\usepackage{enumerate}
\usepackage{slashed}
\usepackage{graphicx}
\usepackage{newlfont}
\usepackage{amsrefs}
\usepackage{comment}
\usepackage[normalem]{ulem}
\usepackage{mathtools}
\usepackage{accents}
\usepackage{leftidx}

\numberwithin{equation}{section}

\theoremstyle{plain}
\newtheorem{theorem}{Theorem}

\theoremstyle{definition}

\theoremstyle{remark}

\newcommand{\overbar}[1]{\mkern 1.6mu\overline{\mkern-1.6mu#1\mkern-1.6mu}\mkern 1.6mu}

\begin{document}

\title[Bekenstein Bounds and Penrose Inequalities, and Black Hole Formation]{Bekenstein Bounds, Penrose Inequalities, and Black Hole Formation}

\author[Jaracz]{Jaroslaw S. Jaracz}
\author[Khuri]{Marcus A. Khuri}
\address{Department of Mathematics\\
Stony Brook University\\
Stony Brook, NY 11794, USA}
\email{jaroslaw.jaracz@stonybrook.edu, khuri@math.sunysb.edu}

\thanks{M. Khuri acknowledges the support of NSF Grant DMS-1708798.}

\begin{abstract}
A universal geometric inequality for bodies relating energy, size, angular momentum, and charge is naturally implied by Bekenstein's entropy bounds. We establish versions of this inequality for axisymmetric bodies satisfying appropriate energy conditions, thus lending credence to the most general form of Bekenstein's bound. Similar techniques are then used to prove a Penrose-like inequality in which the ADM energy is bounded from below in terms of horizon area, angular momentum, and charge. Lastly, new criteria for the formation of black holes is presented involving concentration of angular momentum, charge, and nonelectromagnetic matter energy.
\end{abstract}
\maketitle

\section{Introduction}
\label{sec1} \setcounter{equation}{0}
\setcounter{section}{1}

In \cite{Bekenstein} Bekenstein utilized heuristic arguments involving black holes to derive an upper bound for the entropy of macroscopic bodies, in terms of the total energy and radius of the smallest sphere that encloses the object. This inequality was later generalized \cite{BekensteinMayo,Hod,Zaslavskii} to include contributions from the angular momentum $\mathcal{J}$ and charge $Q$ of the body
\begin{equation}\label{1}
 \sqrt{(\mathcal{E}\mathcal{R})^2 -c^2 \mathcal{J}^2} - \frac{Q^2}{2}
\geq \frac{\hbar c}{2 \pi k_b}\mathcal{S},
\end{equation}
where Boltzmann's constant is denoted by $k_b$, $S$ is entropy, $\mathcal{E}$ is total energy, $\mathcal{R}$ is the radius described above, $\hbar$ and $c$ are the reduced Planck's constant and speed of light. Although the original inequality \cite{Bekenstein} without angular momentum and charge has received much attention \cite{Bekenstein1,Bousso,Wald1},
the enhanced relation \eqref{1} has not been properly investigated. An important initial step in that direction was taken by Dain \cite{Dain} who studied the inequality
\begin{equation}\label{2}
\mathcal{E}^2\geq \frac{Q^4}{4\mathcal{R}^2}+\frac{c^2\mathcal{J}^2}{\mathcal{R}^2},
\end{equation}
which is implied by \eqref{1} since entropy is always nonnegative. He was able to establish \eqref{2} within the context of electromagnetism, and also in general relativity for bodies with zero angular momentum contained in asymptotically flat, maximal initial data which are void of black holes. In this result $\mathcal{E}$ is given by the ADM energy. The idea is that a proof of \eqref{2} lends indirect evidence for the full Bekenstein bound \eqref{1}. Later on Dain's result was extended to include a contribution from angular momentum \cite{AngladaGabach-ClementOrtiz}, again in the setting of asymptotically flat, maximal initial data. The inequality produced in \cite{AngladaGabach-ClementOrtiz} is not quite in the form of \eqref{2}, and it is not clear if one implies the other. Both results \cite{Dain,AngladaGabach-ClementOrtiz} are based on monotonicity of the Hawking mass along inverse mean curvature flow (IMCF), which is valid in the maximal case assuming the dominant energy condition holds.

The purpose of the present article is threefold. The first goal is to establish Bekenstein-like inequalities closely related to \eqref{2} without the hypothesis of maximality for the initial data, and thereby generalize the works \cite{Dain,AngladaGabach-ClementOrtiz}. Our approach relies on a coupling of the IMCF with an embellished version of the Jang equation \cite{BrayKhuri,BrayKhuri1}, which is inspired by the proof of the positive mass theorem \cite{SchoenYau}. Secondly, our techniques naturally lend themselves to establish a version of the Penrose inequality \cite{Mars} with angular momentum and charge, for general axisymmetric initial data without the maximal assumption. A similar result in the maximal case was recently given in \cite{Anglada}. Recall that Penrose \cite{Penrose} proposed a sharp inequality bounding the total energy of a black hole spacetime from below in terms of the horizon area. It serves a necessary condition for the cosmic censorship conjecture. Thus, a counterexample would disprove cosmic censorship while verification of the Penrose inequality only lends credence to the conjecture's validity. In \cite{Bray,HuiskenIlmanen} the Penrose inequality has been proven in the maximal case. Moreover, generalizations including angular momentum and charge have been proposed \cite{Mars} motivated by Penrose's original heuristic arguments. The full Penrose inequality may then be stated as follows
\begin{equation}\label{3}
\mathcal{E}^2\geq\left(\frac{c^4}{G}\sqrt{\frac{A}{16\pi}}+\sqrt{\frac{\pi}{A}}Q^2\right)^2
+\frac{4\pi c^2\mathcal{J}^2}{A},
\end{equation}
where $A$ is the minimum area required to enclose the outermost apparent horizon in an axisymmetric initial data set satisfying the relevant energy condition, and $G$ is the gravitational constant. This comes with a rigidity statement asserting that equality holds if and only if the initial data arise from an embedding into the Kerr spacetime. We also note that the Bekenstein bound \eqref{1}, when applied to black holes, implies the Penrose inequality \eqref{3}. To see this, simply recall that for a black hole with event horizon area $A_e$ the radius and entropy are given by
\begin{equation}\label{4}
\mathcal{R}=\sqrt{\frac{A_e}{4\pi}},\quad\quad\quad \mathcal{S}=\frac{k_{b}A_{e}}{4l_{p}^2},
\end{equation}
where $l_p=\sqrt{G\hbar/c^3}$ is the Planck length. Inequality \eqref{3} has been established in the maximal case without the angular momentum term in a series of papers \cite{DisconziKhuri,KhuriWeinsteinYamada,KhuriWeinsteinYamada1,KhuriWeinsteinYamada2}.
However, there has been very little to no progress on including angular momentum. The only result known to the author in this direction is \cite{Anglada}. Here we will establish a version of \eqref{3} valid in the general nonmaximal setting, assuming the existence of solutions to a canonical coupling of the Jang equation to IMCF; such solutions are known to exist in spherical symmetry.

Lastly, the methods used to study the Penrose inequality above lead to new criteria for black hole formation, as well as inequalities for bodies involving size, angular momentum, and charge. Recall that Thorne's hoop conjecture \cite{Thorne} roughly states that if enough matter/energy is condensed in an appropriately small region, then gravitational collapse will ensue. Mathematically this assertion may be translated into a heuristic inequality
\begin{equation}\label{5}
\mathrm{Mass}(\Omega)>\mathcal{C}\cdot\mathrm{Size}(\Omega),
\end{equation}
which if satisfied for a body $\Omega$, then implies that $\Omega$ must be contained within an apparent horizon; here $\mathcal{C}$ is a universal constant. One of the primary difficulties in establishing such a result is finding a proper notion of quasi-local mass to use in the left-hand side of \eqref{5}. As it turns out, mass/energy is not the only quantity that is appropriate to place on the left-hand side of the inequality. We will show below that angular momentum and charge also naturally arise on the left-hand side, and thus provide extra means to satisfy \eqref{5}. This will be rigorously proven in spherical symmetry, and motivation will be given to indicate why the result should hold in generality. Related results concerning black hole existence due to concentration of angular momentum or charge have been given in \cite{Khuri,Khuri1,KhuriXie}, using different methods.

\section{Bekenstein Bounds}
\label{sec2} \setcounter{equation}{0}
\setcounter{section}{2}

Consider a spacelike slice $M$ of an asymptotically flat 4-dimensional spacetime. The induced positive definite metric $g$ and extrinsic curvature $k$ together yield an initial data set $(M,g,k)$. If $T_{ab}$ denotes the stress energy tensor and $n^a$ is the unit timelike normal to $M$, then $\mu=T_{ab}n^a n^b$ and $c^{-1}J_i=c^{-1}T_{ai}n^a$ are the matter energy and momentum density of the slice. These must satisfy the constraint equations
\begin{align}\label{2.1}
\begin{split}
\frac{16\pi G}{c^{4}}\mu &= R+(Tr_{g}k)^{2}-|k|^{2},\\
\frac{8\pi G}{c^{4}} J &= \operatorname{div}(k-(Tr_{g}k)g),
\end{split}
\end{align}
where $R$ is the scalar curvature of $g$. We will also be interested in the electromagnetic field, and let $E$ and $B$ denote the electric and magnetic fields induced on $M$, respectively. Assuming that all measured energy densities are nonnegative implies $\mu\geq |J|$, which is referred to as the dominant energy condition. It is also useful to single out the nonelectromagnetic matter fields for which the energy and momentum densities are obtained from $\mu_{EM} = \mu-\frac{1}{8\pi}\left(|E|^{2}+|B|^{2}\right)$ and $J_{EM} = J +\frac{1}{4\pi}E\times B$. The charged dominant energy condition is then $\mu_{EM}\geq|J_{EM}|$.

A body $\Omega$ will be described as a connected open subset of $M$ having compact closure and smooth boundary $\partial\Omega$. The total charge within the body is then given by
\begin{equation}\label{2.2}
Q^2=\left(\frac{1}{4\pi}\int_{\Omega}\operatorname{div} E dx_g\right)^2
+\left(\frac{1}{4\pi}\int_{\Omega}\operatorname{div} B dx_g\right)^2,
\end{equation}
and it will always be presumed that there is no charged matter outside $\Omega$. In order to characterize the angular momentum of the body, the initial data will be assumed to be axisymmetric. That is, there is a $U(1)$ subgroup within the group of isometries of the Riemannian manifold $(M, g)$, and all relevant quantities are invariant under the $U(1)$ action. Without axisymmetry it is problematic to define quasi-local angular momentum \cite{Szabados}. Moreover, with this hypothesis all angular momentum is contained within the matter fields, as gravitational waves carry no angular momentum.
Let $\eta$ be the generator of the $U(1)$ symmetry, then the angular momentum of the body is
\begin{equation}\label{2.3}
\mathcal{J}=\frac{1}{c}\int_{\Omega}J_i \eta^i dx_g.
\end{equation}

The basic strategy to obtain Bekenstein type bounds \eqref{2} is to use monotonicity of the Hawking mass along inverse mean curvature flow. This worked well in \cite{Anglada,AngladaGabach-ClementOrtiz,Dain} because of the maximal assumption $Tr_{g}k=0$. More precisely, monotonicity of the Hawking mass relies on nonnegativity of the scalar curvature, and this is achieved with the dominant energy condition if the data are maximal. Here we do not assume that the data are maximal, and thus this method breaks down. However, we may follow an approach similar to that in the proof of the positive mass theorem \cite{SchoenYau}, where the initial data are deformed by $(M,g,k)\rightarrow (M,\overbar{g})$ with $\overbar{g}_{ij}=g_{ij}+u^2 f_i f_j$ for some functions $u>0$ and $f$. In \cite{SchoenYau} the function $u=1$ and $f$ is chosen to solve the so called Jang equation, which is designed to impart positivity properties to the scalar curvature $\overbar{R}$ of $\overbar{g}$. In the present setting it is more appropriate to utilize an embellished version of the Jang equation
\begin{equation}\label{2.4}
\left(g^{ij}-\frac{u^2 f^{i}f^{j}}{1+u^2|\nabla f|^{2}}\right)
\left(\frac{u\nabla_{ij}f+u_i f_j +u_j f_i}
{\sqrt{1+u^2 |\nabla f|^{2}}}-k_{ij}\right)=0,
\end{equation}
where $\nabla_{ij}$ are second covariant derivatives with respect to $g$ and $f^i=g^{ij}f_{j}$. This equation also yields desirable features for the scalar curvature which now takes the form
\begin{equation}\label{2.5}
\overbar{R}=\frac{16\pi G}{c^4}(\mu-J(w))+
|h-k|_{\overbar{g}}^{2}+2|q|_{\overbar{g}}^{2}
-2u^{-1}\operatorname{div}_{\overbar{g}}(uq),
\end{equation}
where
\begin{equation}\label{2.6}
h_{ij}=\frac{ u \nabla_{ij}f
+ u_i f_j +  u_j  f_i}{ \sqrt{1 + u^2 |\nabla f|^2 }},\text{ }\text{ }\text{ }\text{ }
w_{i}=\frac{u f_{i}}{\sqrt{1+u^{2}|\nabla f|^{2}}},\text{
}\text{ }\text{ }\text{ }
q_{i}=\frac{u f^{j}}{\sqrt{1+u^{2}|\nabla f|^{2}}}(h_{ij}-k_{ij}).
\end{equation}
These formulas along with their geometric interpretations are explained in \cite{BrayKhuri,BrayKhuri1}. Observe that the first term on the right-hand side of \eqref{2.5} is nonnegative if the dominant energy condition is satisfied, since $|w|\leq 1$. Furthermore, all other terms are manifestly nonnegative except possibly the divergence term. The deformed scalar curvature may then be described as `weakly' nonnegative, since integrating it against $u$ produces a nonnegative quantity modulo boundary terms.

In order to optimize the positivity of $\overbar{R}$ with regards to IMCF we choose $u$ as follows. Let $\{\overline{S}_{t}\}_{t=t_0}^{\infty}$ be an IMCF in the deformed data $(M,\overbar{g})$, where $t_0=0$ or $-\infty$ depending on whether the flow starts at a surface or a point. A weak version of the flow always exists \cite{HuiskenIlmanen} in the asymptotically flat setting, although for the purposes of exposition we may assume that the flow is smooth. Then set $u=\sqrt{|\overbar{S}_t|/16\pi}\text{ }\!\overbar{H}$ to be the product of the square root of area and mean curvature for the flow surfaces. Consider now the Hawking energy of the flow surfaces within the deformed data
\begin{equation}\label{2.7}
\mathcal{E}_{H}(\overbar{S}_{t})=\frac{c^4}{G}\sqrt{\frac{|\overbar{S}_t|}{16\pi}}
\left(1-\frac{1}{16\pi}\int_{\overbar{S}_t}\overbar{H}^2\right).
\end{equation}
A well-known computation \cite{HuiskenIlmanen} asserts that if $t_2>t_1$ then
\begin{equation}\label{2.8}
\mathcal{E}_{H}(\overbar{S}_{t_2})-\mathcal{E}_{H}(\overbar{S}_{t_1})
\geq \frac{c^4}{16\pi G}\int_{t_1}^{t_2}\sqrt{\frac{|\overbar{S}_t|}{16\pi}}
\int_{\overbar{S}_t}\overbar{R}.
\end{equation}
The first two terms on the right-hand side in the expression \eqref{2.5} will provide lower bounds for \eqref{2.8} involving the charge and angular momentum, while the divergence expression will contribute to the Hawking energies.

Consider now the case when the flow starts from a point $x_0$ within the body $\Omega$ on the axis of rotation, so that the starting time is $t_0=-\infty$. Observe that in \eqref{2.8} with $t_1=-\infty$ and $t_2=\infty$ several simplifications occur. Namely, since the Hawking energy of a point is zero and the limit of Hawking energies as $t\rightarrow\infty$ is no larger than the ADM (total) energy, the left-hand side of \eqref{2.8} may be replaced with the ADM energy $\mathcal{E}$. Note that this total energy is a priori with respect to the deformed metric $\overbar{g}$. However by placing the natural boundary conditions at infinity for solutions of the Jang equation, namely $f\rightarrow 0$ in the asymptotic end, the total energy of $g$ and $\overbar{g}$ are equivalent \cite{SchoenYau}. Furthermore if the charged dominant energy condition holds
then $\mu-J(w)\geq \frac{1}{8\pi}(|E|^2+|B|^2)$, as it may be assumed without loss of generality in axisymmetry that the electric and magnetic fields have no component in the Killing direction so that $E\times B(w)=0$. In \cite{DisconziKhuri} a deformation of the electromagnetic field $(E,B)\rightarrow (\overbar{E},\overbar{B})$, tailored to the Jang metric $\overbar{g}$, was given which preserves total charge as well as zero charge density and has less energy density than the original field $|E|\geq|\overbar{E}|$, $|B|\geq|\overbar{B}|$. From this a lower bound for the right-hand side of \eqref{2.8} is obtained in terms of the energy density of $(\overbar{E},\overbar{B})$, and since the surface integrals are computed with respect to $\overbar{g}$ a relation with total charge is produced as in \cite{DisconziKhuri}. In particular
\begin{equation}\label{2.9}
\int_{-\infty}^{\infty}\sqrt{\frac{|\overbar{S}_t|}{16\pi}}\int_{\overbar{S}_t}(\mu-J(w))
\geq\frac{1}{8\pi}\int_{t_*}^{\infty}\sqrt{\frac{|\overbar{S}_t|}{16\pi}}
\int_{\overbar{S}_t}(|\overbar{E}|^2+|\overbar{B}|^2)
\geq\frac{Q^2}{2\overbar{\mathcal{R}}_{t_*}},
\end{equation}
where $\overbar{\mathcal{R}}_{t_*}=\sqrt{|\overline{S}_{t_*}|/4\pi}$ is the area radius of $\overbar{S}_{t_*}$. The time $t_*$ may be chosen arbitrarily, however in order to obtain the optimal inequality for the body, $t_*$ will denote the first (smallest) time such that the flow surface $\overbar{S}_{t_*}$ completely encloses $\Omega$. Moreover since the flow will change depending on the choice of its starting point $x_0$, optimization requires that we choose the $x_0$ for which the area radius at $t_*$ is smallest. Such a starting point exists within the body since $\Omega$ is compact. The radius $\overbar{\mathcal{R}}$ of $\Omega$ will then be defined as in \cite{Dain} to be this optimal area radius, and in \eqref{2.9} the radius $\overbar{\mathcal{R}}_{t_*}$ may be replaced with $\overbar{\mathcal{R}}$.

Within the scalar curvature formula \eqref{2.5} the second term on the right-hand side encodes a contribution from angular momentum. In order to extract this contribution we first make some observations. The metric $\overbar{g}$ arises as the induced metric on the graph of the function $f$ \cite{BrayKhuri1}, and the surfaces $\overbar{S}_t$
may be interpreted as a flow within the graph. There is then a natural projection of $\overbar{S}_t$ into $(M,g)$ which will be denoted $S_t$. Since the flow starts from a point on the symmetry axis, each of the surfaces $\overbar{S}_t$, $S_t$ is axisymmetric. As is shown in the appendix under mild hypotheses, it then follows that $h(\eta,\nu)=0$ on $S_t$, where $\nu$ is the unit normal to $S_t$. Therefore assuming that angular momentum density vanishes outside the body and using H\"{o}lder's inequality produces
\begin{align}\label{2.10}
\begin{split}
\left(\frac{8\pi G}{c^3}\right)^2\mathcal{J}^2=\left(\int_{S_t}k(\eta,\nu)\right)^2
= &\left(\int_{S_t}[k(\eta,\nu)-h(\eta,\nu)]\right)^2\\
\leq &\left(\int_{S_t}|k-h|_{g}|\eta|\right)^2
\leq\int_{\overbar{S}_t}|k-h|_{\overbar{g}}^2 \int_{\overbar{S}_t}|\eta|^2,
\end{split}
\end{align}
where we have also used the fact that $\overbar{g}$ measures areas to be at least as large as does $g$. This estimate is suited to give a lower bound for the ADM energy which may be expressed properly with the `circumference' radius
\begin{equation}\label{2.11}
\overbar{\mathcal{R}}_{c}^{\text{ }\!-2}=\sqrt{|\overbar{S}_{t_*}|}
\int_{t_*}^{\infty}\frac{\sqrt{|\overbar{S}_t|}}{\int_{\overbar{S}_t}|\eta|^2}.
\end{equation}
The radius $\overbar{\mathcal{R}}_{c}$ was used and studied in \cite{Anglada,AngladaGabach-ClementOrtiz}, where it was shown that if the flow has reasonably nice properties then this radius may be related to more traditional measures of size for the body. In particular if the flow remains convex outside of $\Omega$, as it is known to be for large times $|t|>>0$ or in spherical symmetry, then $\overbar{\mathcal{R}}_{c}\leq \sqrt{5/2}\max_{S_{t_*}}|\eta|$ which is proportional to the circumference of the largest orbit within $S_{t_*}$. Because it provides and upper bound for $\overbar{\mathcal{R}}_{c}$, when the flow is convex the circumference may be used in place of the this radius in
\begin{equation}\label{2.12}
\frac{c^{4}}{16\pi G}\int_{0}^{\infty}\sqrt{\frac{|\overbar{S}_{t}|}{16\pi}}
\int_{\overbar{S}_{t}}|h-k|_{\overline{g}}^2
\geq \frac{G}{2c^2}\frac{\mathcal{J}^2}{\overbar{\mathcal{R}}
\text{ }\!\overbar{\mathcal{R}}^{2}_{c}}.
\end{equation}

It is now possible to combine \eqref{2.8}, \eqref{2.9}, and \eqref{2.12} to obtain a Bekenstein-type bound. Note that the proof above relies on the existence of a solution to the Jang equation coupled to IMCF through the choice of the function $u$. Due to the fact that solutions to the Jang equation tend to blow-up at apparent horizons \cite{HanKhuri}, it will be assumed that the initial data are devoid of these surfaces.
Under this hypothesis, the desired solutions to the Jang/IMCF system have been shown to always exist in spherical symmetry \cite{BrayKhuri}, and it is reasonable to expect that existence continues to hold at least in a weak sense in axisymmetry.

\begin{theorem}\label{thm1}
Let $(M,g,k,E,B)$ be a complete, axisymmetric, asymptotically flat initial data set for the Einstein-Maxwell equations, satisfying the charged dominant energy condition $\mu_{EM}\geq|J_{EM}|$ and without apparent horizons. Suppose that $\Omega\subset M$ is a body outside of which there is no charge density or momentum density in the direction of axisymmetry. If the Jang/IMCF system of equations admits a solution then
\begin{equation}\label{2.13}
\mathcal{E}\geq \frac{Q^2}{2\overbar{\mathcal{R}}}+\frac{G}{2c^2}\frac{\mathcal{J}^2}{\overbar{\mathcal{R}}
\text{ }\!\overbar{\mathcal{R}}^{2}_{c}}.
\end{equation}
\end{theorem}

This theorem generalizes the results of \cite{AngladaGabach-ClementOrtiz,Dain} to the non-maximal setting. Although it is in the spirit of the Bekenstein bound \eqref{2}, these two inequalities are distinct in that one does not directly imply the other. Nevertheless, as will be shown in the next section inequality \eqref{2.13} does indirectly imply a lower bound for $\mathcal{E}^2$ which has the same structure as \eqref{2}.

\section{Penrose Inequalities}
\label{sec3} \setcounter{equation}{0}
\setcounter{section}{3}

In this section we will adapt the techniques discussed above to establish a version of the Penrose inequality with angular momentum and charge \eqref{3}. This will then yield an alternate version of the Bekenstein bound \eqref{2.13}. Recall that an apparent horizon is a surface $S\subset M$ which has zero null expansion, that is, a shell of light emitted from the surface is (infinitesimally) neither growing nor shrinking in area as it leaves the surface. These surfaces indicate the presence of a strong gravitational field, and may be interpreted as quasi-local versions of black hole event horizons from the initial data point of view. Mathematically they are expressed by one of the two equations $\theta_{\pm}:=H\pm Tr_{S}k=0$, where the signs $+/-$ indicate a future/past horizon. An apparent horizon is called outermost within an initial data set if it is not enclosed by any other apparent horizon.

In contrast to the previous section, here we will work with an IMCF starting at a closed axisymmetric surface $S$ so that $t_0=0$ is the starting time of the flow, and $S$ will either be an outermost apparent horizon or the boundary of a body $\partial\Omega$. First consider the case in which $S=\partial\Omega$, and assume that the boundary of the body is completely untrapped $H>|Tr_{S}k|$. This allows for the prescription of a Neumann type boundary condition for solutions of the Jang equation \eqref{2.4}
\begin{equation}\label{3.1}
\frac{u\partial_{\nu}f}{\sqrt{1+u^2 |\nabla f|^2}}=H^{-1}Tr_{S}k.
\end{equation}
It was shown in \cite{Khuri0,BrayKhuri}, in the context of spherical symmetry, that solutions of the Jang/IMCF system exist satisfying this boundary condition.
Moreover it was also shown that with \eqref{3.1} the boundary integrals arising from the divergence expression associated with $\overbar{R}$ in \eqref{2.8}, combine with the Hawking energy on the left-hand side of \eqref{2.8}, to yield
\begin{equation}\label{3.2}
\mathcal{E}-\mathcal{E}_{SH}(S)\geq
\int_{0}^{\infty}\sqrt{\frac{|\overbar{S}_t|}{16\pi}}
\int_{\overbar{S}_t}\left((\mu-J(w))+\frac{c^4}{16\pi G}|h-k|_{\overbar{g}}^2\right)
\end{equation}
where the spacetime Hawking energy is given by
\begin{equation}\label{3.3}
\mathcal{E}_{SH}(S)=\frac{c^4}{G}\sqrt{\frac{|S|}{16\pi}}\left(1-\frac{1}{16\pi}
\int_{S}\theta_{+}\theta_{-}\right).
\end{equation}
It should be pointed out that \eqref{3.2} depends on appropriate behavior of the IMCF. For instance in the weak formulation of Huisken/Ilmanen \cite{HuiskenIlmanen}, the flow may instantaneously jump from the desired starting surface $S$ to another surface $\tilde{S}$ enclosing it with less area. If this occurs, then in inequality \eqref{3.2} the role of $S$ should be replaced by $\tilde{S}$. Such `jumping' behavior can be prevented by requiring that $S$ be outer area minimizing in $(M,\overbar{g})$, in that any surface which encloses $S$ should have greater area. In order to achieve this property with respect to the deformed data metric $\overbar{g}$, further geometric hypotheses on $S$ with respect to the original initial data may be required. For the purposes of the present article, which does not seek to fully examine the analytical problem of solving the Jang/IMCF system in generality, we will simply refer to solutions with these suitable properties as \textit{proper solutions}. As pointed out, it is known that proper solutions always exist under the hypothesis of spherical symmetry and small perturbations thereof.

As in the previous section, the two terms on the right-hand side of \eqref{3.2} yield contributions of angular momentum and charge. More precisely, applying \eqref{2.9} and \eqref{2.12} produces
\begin{equation}\label{3.4}
\mathcal{E}\geq \mathcal{E}_{SH}(S)+\frac{Q^2}{2\overbar{\mathcal{R}}_0}
+\frac{G}{2c^2}\frac{\mathcal{J}^2}{\overbar{\mathcal{R}}_0\overbar{\mathcal{R}}_{c}^2}
\end{equation}
where $\overbar{\mathcal{R}}_0$ is the area radius of $S_0=S$, and $\mathcal{J}$, $Q$ denote the angular momentum and charge contained within $S$. This inequality will lead to a Bekenstein bound for bodies in the presence of a sufficiently strong gravitation field, as well as a version of the Penrose inequality.

The arguments above seem to rely on the assumption that $S$ is untrapped, as otherwise the boundary condition \eqref{3.1} would imply that $u\partial_{\nu}f=\pm \infty$. However for the Jang equation, blow-up solutions are natural as first observed in the proof of the positive mass theorem \cite{SchoenYau}. Blow-up occurs at apparent horizons, and can be prescribed at outermost apparent horizons as well \cite{HanKhuri}. Therefore in place of the boundary condition \eqref{3.1}, at an outermost apparent horizon $S$ we will prescribe blow-up as the boundary condition. In this situation the graph of the solution to Jang's equation asymptotes to a cylinder over $S$, and the area of this surface in the deformed metric and the original coincide $|\overbar{S}|=|S|$. Moreover, at an apparent horizon $\theta_{+}\theta_{-}=0$ so that $\mathcal{E}_{SH}(S)=\sqrt{|S|/16\pi}$.
With this a version of the Penrose inequality follows.

\begin{theorem}\label{thm2}
Let $(M,g,k,E,B)$ be an axisymmetric, asymptotically flat initial data set for the Einstein-Maxwell equations, satisfying the charged dominant energy condition $\mu_{EM}\geq|J_{EM}|$ and with outermost apparent horizon boundary having one component.
Suppose further that there is no charge density or momentum density in the direction of axisymmetry. If the Jang/IMCF system of equations admits a proper solution then
\begin{equation}\label{3.5}
\mathcal{E}^2\geq \left(\frac{c^4}{G}\sqrt{\frac{|\partial M|}{16\pi}}+\sqrt{\frac{\pi}{|\partial M|}}Q^2\right)^2 +\frac{c^2\mathcal{J}^2}{4\overbar{\mathcal{R}}^{2}_{c}}.
\end{equation}
\end{theorem}

This result is similar to the conjectured Penrose inequality \eqref{3} with the primary difference arising in the angular momentum term. Instead of area, this term involves the squared radius defined in the previous section. The proof of this theorem requires one more observation in order that it follow from \eqref{3.4}. Namely, upon multiplying \eqref{3.4} by the first two terms on the right-hand side we find
\begin{equation}\label{3.6}
\mathcal{E}^2\geq\left(\mathcal{E}_{SH}(S)+\frac{Q^2}{2\overbar{\mathcal{R}}_0}\right)
\mathcal{E}
\geq \left(\mathcal{E}_{SH}(S)+\frac{Q^2}{2\overbar{\mathcal{R}}_0}\right)^2
+\mathcal{E}_{SH}(S)\frac{G}{2c^2}
\frac{\mathcal{J}^2}{\overbar{\mathcal{R}}_0\overbar{\mathcal{R}}_{c}^2}.
\end{equation}
From this the desired inequality in Theorem \ref{thm2} arises from the arguments above. Furthermore \eqref{3.6} may be used to yield a Bekenstein bound. Suppose that $S=\partial\Omega$ is the boundary of a body immersed in a strong gravitational field. By this we mean that $\lambda:=1-(|S|/16\pi) \sup_{S}\theta_{+}\theta_{-}>0$, or rather that $\theta_{+}\theta_{-}$ has sufficiently small positive part. In particular surfaces $S$ which are close to being an apparent horizon satisfy this property, as do trapped surfaces. For surfaces $S$ which satisfy this property, the spacetime Hawking energy is bounded below by the product of $\lambda$ and the area radius up to a universal constant. Let $\lambda_0>0$ be fixed and consider the class of bodies with boundaries experiencing a appropriately strong gravitational field so that $\lambda\geq \lambda_0$. Then for bodies of this type a Bekenstein bound follows immediately from \eqref{3.6}
\begin{equation}\label{3.7}
\mathcal{E}^2\geq \frac{Q^4}{4\overbar{\mathcal{R}}_{0}^2}
+\lambda_0 \frac{c^2\mathcal{J}^2}{4\overbar{\mathcal{R}}^{2}_{c}}.
\end{equation}
This inequality has the same structure as the Bekenstein inequality \eqref{2}, although the radius associated with the angular momentum term is more complicated than the area radius.

\section{Black Hole Formation}
\label{sec4} \setcounter{equation}{0}
\setcounter{section}{4}

It is a basic folklore belief that if enough matter/energy is concentrated in a sufficiently small region, then gravitational collapse must ensue. This is typically
referred to as the hoop conjecture or trapped surface conjecture \cite{Seifert,Thorne}, and is quite difficult to formulate precisely, see the references in \cite{KhuriXie}.
One of the most general results in this direction is due to Schoen and Yau \cite{SchoenYau1}, who exploited the techniques developed in their proof of the positive mass theorem \cite{SchoenYau} to prove the existence of apparent horizons whenever matter density is highly concentrated. Their strategy is to show that the concentration hypothesis forces solutions of the Jang equation to blow-up, and since blow-up can only occur at an apparent horizon the existence of such a surface in the initial data is established. Here we will combine this strategy with the techniques used in the previous two sections to obtain a black hole existence result due to concentration of nonelectromagnetic matter energy, charge, or angular momentum. In addition, the measure of size used in our result will differ considerably from the complicated measure in \cite{SchoenYau}.

Consider two concentric bodies $\Omega_1\subset\Omega_2$, each having the topology of a 3-dimensional ball, inside an axisymmetric asymptotically flat initial data set $(M,g,k,E,B)$. The model astronomical body in this context is a typical star, where there is a highly dense core and interior (represented by $\Omega_1$) compared to the outermost layer or corona (represented by $\Omega_2\setminus\Omega_1$) with very little matter density. For simplicity of the model we will assume that the charge density and momentum density in the Killing direction vanish in the annular region $\Omega_2\setminus\Omega_1$, so that $\operatorname{div}E=\operatorname{div}B=J(\eta)=0$. If there are no apparent horizons in the initial data, then as discussed in Section \ref{sec2} we may take a solution of the Jang/IMCF system of equations with the flow emanating from a point $x_0\in\Omega_1$ on the axis of rotation. Let $t_1$ and $t_2$ be the first times for which the flow completely encloses the boundaries $\partial\Omega_1$ and $\partial\Omega_2$, respectively. From the arguments used to obtain \eqref{3.2}, together with the fact that the Hawking energy of a point is zero, we find that
\begin{equation}\label{4.1}
\mathcal{E}_{SH}(S_{t_2})\geq
\int_{-\infty}^{t_2}\sqrt{\frac{|\overbar{S}_t|}{16\pi}}
\int_{\overbar{S}_t}\left((\mu-J(w))+\frac{c^4}{16\pi G}|h-k|_{\overbar{g}}^2\right)
\end{equation}
if the Jang solution $f$ is prescribed to be zero (or more generally constant) on $S_{t_2}$. Note that this boundary condition differs from \eqref{3.1} which is used to obtain \eqref{3.2}. This is due to the fact that the boundary integrals that arise from the divergence term in $\overbar{R}$ have different signs on the inner and outer boundaries \cite{Khuri0}. In fact the boundary terms at the outer boundary have an advantageous sign, and it is likely that this Dirichlet boundary condition used for \eqref{4.1} is not needed.

Proceeding as in Section \ref{sec2}, lower bounds for the right-hand side of \eqref{4.1} may be extracted in terms of the total charge and angular momentum of $\Omega_1$. In addition, a contribution from the nonelectromagnetic matter fields will also occur. To see this observe that as in \eqref{2.9}
\begin{align}\label{4.2}
\begin{split}
\int_{-\infty}^{t_2}\sqrt{\frac{|\overbar{S}_t|}{16\pi}}
\int_{\overbar{S}_t}(\mu-J(w))\geq&\int_{-\infty}^{t_1}\sqrt{\frac{|\overbar{S}_t|}{16\pi}}
\int_{\overbar{S}_t}(\mu_{EM}-J_{EM}(w))+\frac{1}{8\pi}\int_{t_1}^{t_2}\sqrt{\frac{|\overbar{S}_t|}{16\pi}}
\int_{\overbar{S}_t}\left(|\overbar{E}|^2+|\overbar{B}|^2\right)\\
\geq&\frac{4\pi}{3}\overbar{\mathcal{R}}_{1}^3
\min_{\tilde{\Omega}_{1}}(\mu_{EM}-|J_{EM}|)+\frac{Q^2}{2\overbar{\mathcal{R}}_{1}}
\left(1-\sqrt{\frac{\overbar{\mathcal{R}}_{1}}{\overbar{\mathcal{R}}_{2}}}\right),
\end{split}
\end{align}
where $\tilde{\Omega}_{1}$ is the domain enclosed by $\overbar{S}_{t_1}$ and $\overbar{\mathcal{R}}_{1}$, $\overbar{\mathcal{R}}_{2}$ are the area radii of $\overbar{S}_{t_1}$, $\overbar{S}_{t_2}$. Notice that if the charged dominant energy condition is valid then the first term on the right in \eqref{4.2} is nonnegative, and the second term also has this property since areas are nondecreasing in an IMCF. Similarly, applying the arguments of \eqref{2.12} to the current setting produces
\begin{equation}\label{4.3}
\frac{c^4}{16\pi G}\int_{-\infty}^{t_2}\sqrt{\frac{|\overbar{S}_t|}{16\pi}}
\int_{\overbar{S}_t}|h-k|_{\overbar{g}}^2
\geq \frac{G}{2c^2}\frac{\mathcal{J}^2}{\overbar{\mathcal{R}}_1 \overbar{\mathcal{R}}_{ac}^2},
\end{equation}
where the circumference radius is with respect to the annular domain
\begin{equation}\label{4.4}
\overbar{\mathcal{R}}_{ac}^{-2}=\sqrt{|\overbar{S}_{t_1}|}
\int_{t_1}^{t_2}\frac{\sqrt{|\overbar{S}_t|}}{\int_{\overbar{S}_t}|\eta|^2}.
\end{equation}
Furthermore assuming that the outer surface $S_{t_2}$ is untrapped, so that $H>|Tr_{S_{t_2}}k|$, implies that the Hawking energy may be estimated above by the area radius $\mathcal{E}_{SH}(S_{t_2})\leq\frac{c^4}{2G}\overbar{\mathcal{R}}_{2}$. Therefore combining \eqref{4.1}, \eqref{4.2}, and \eqref{4.3} yields
\begin{equation}\label{4.5}
\frac{c^4}{2G}\overbar{\mathcal{R}}_{2}
\geq\frac{4\pi}{3}\overbar{\mathcal{R}}_{1}^3
\min_{\tilde{\Omega}_{1}}(\mu_{EM}-|J_{EM}|)+\frac{Q^2}{2\overbar{\mathcal{R}}_{1}}
\left(1-\sqrt{\frac{\overbar{\mathcal{R}}_{1}}{\overbar{\mathcal{R}}_{2}}}\right)
+\frac{G}{2c^2}\frac{\mathcal{J}^2}{\overbar{\mathcal{R}}_1\overbar{\mathcal{R}}_{ac}^2}.
\end{equation}

The geometric inequality \eqref{4.5} relates the size of the body $\Omega_2\supset\Omega_1$ to its core nonelectromagnetic matter content, total charge, and total angular momentum. It may be interpreted as stating that a material body of fixed size can only contain a certain fixed amount of matter energy, charge, and angular momentum. The primary hypotheses which were used to derive this inequality consist of the assumption that the outer region is untrapped, the annular region $\Omega_{2}\setminus\Omega_1$ has no charge and momentum density in the Killing direction, and most importantly that the initial data are void of apparent horizons. This latter assumption is used to obtain regular solutions of the Jang equation, and following \cite{SchoenYau1} we may turn this around to obtain a black hole existence result. More precisely if a body with the hypotheses above, minus any assumption on apparent horizons,
satisfies
\begin{equation}\label{4.6}
\frac{c^4}{2G}\overbar{\mathcal{R}}_{2}
<\frac{4\pi}{3}\overbar{\mathcal{R}}_{1}^3
\min_{\tilde{\Omega}_{1}}(\mu_{EM}-|J_{EM}|)+\frac{Q^2}{2\overbar{\mathcal{R}}_{1}}
\left(1-\sqrt{\frac{\overbar{\mathcal{R}}_{1}}{\overbar{\mathcal{R}}_{2}}}\right)
+\frac{G}{2c^2}\frac{\mathcal{J}^2}{\overbar{\mathcal{R}}_1\overbar{\mathcal{R}}_{ac}^2}
\end{equation}
then an apparent horizon must be present within the initial data. The reasoning is that if there were no apparent horizons, then we may apply the arguments above to conclude that \eqref{4.5} holds, a contradiction. This relies on the analysis of the Jang/IMCF system of equations, which has been established rigorously in the case of spherical symmetry \cite{BrayKhuri}. This conclusion concerning the existence of an apparent horizon implies that the spacetime arising from the initial data contains a singularity, or
more accurately is null geodesically incomplete by the Hawking-Penrose singularity theorems \cite{HawkingEllis}, and assuming cosmic censorship it must therefore possess a black hole. This result may be interpreted as stating that if a body of fixed size contains sufficient amounts of nonelectromagnetic matter energy, charge, or angular momentum, then it must collapse to form a black hole.

\appendix
\numberwithin{equation}{section}

\section{Vanishing Extrinsic Curvature}\label{sec5}

Consider an axisymmetric closed surface $S$ within an axisymmetric Riemannian 3-manifold $(M,g)$. If $\eta$ is the generator for the axisymmetry and $\nu$ is the unit normal to $S$, then under mild hypotheses $h(\eta,\nu)=0$ along $S$, where $h$ is the tensor associated with the solution $f$ of Jang's equation and is given in \eqref{2.6}. Geometrically the tensor $h$ represents the extrinsic curvature of the graph of $f$ in a static spacetime constructed from the metric $\overbar{g}$ and function $u$ \cite{BrayKhuri1}. The vanishing of this particular component of $h$ is used throughout the main body of the paper in order to allow for angular momentum contributions to the various inequalities. Here we will confirm this property of $h$.

In \cite{Chrusciel} it was shown that if $M$ is asymptotically flat and simply connected then a global cylindrical coordinate system exists, denoted by $(\rho,z,\phi)$ and referred to as Brill coordinates, such that the metric takes the following form
\begin{equation}\label{5.1}
g=e^{-2U+2\alpha}(d\rho^2 + dz^2)+\rho^2 e^{-2U}(d\phi+A d\rho+B dz)^2
\end{equation}
for some functions $U$, $\alpha$, $A$, and $B$ all depending only on $(\rho,z)$.
The Killing field is given by $\eta=\partial_{\phi}$, and if $U=\alpha=A=B=0$ then
$g$ reduces to the typical expression of the flat metric on Euclidean 3-space in cylindrical coordinates. For simplicity it will be assumed that $A=B=0$ so that $\eta$ is perpendicular to the orbit space or $\rho z$-half plane. Observe that since $u$ and $f$ are axisymmetric, that is $\partial_{\phi}f=\partial_{\phi}u=0$, it follows that
\begin{equation}\label{5.2}
h(\eta,\nu)=\frac{u\nabla_{\phi\nu}f}{\sqrt{1+u^2|\nabla f|^2}}
=-\frac{u\left(\Gamma_{\phi\nu}^{\rho}\partial_{\rho}f
+\Gamma_{\phi\nu}^{z}\partial_{z}f\right)}{\sqrt{1+u^2|\nabla f|^2}}
\end{equation}
where the $\Gamma_{ij}^l$ are Christoffel symbols.  Since the surface is axisymmetric $\partial_{\phi}$ is tangent to $S$, and thus $g(\eta,\nu)=0$.
A straightforward calculation then yields
\begin{equation}\label{5.3}
\Gamma_{\phi\nu}^{\rho}=\frac{1}{2}g^{\rho i}\partial_{\nu} g_{\phi i} =0,\quad\quad\quad
\Gamma_{\phi\nu}^{z}=\frac{1}{2}g^{z i}\partial_{\nu} g_{\phi i} =0,
\end{equation}
and the desired conclusion follows.

\end{document}